\newif\ifpnas
\pnasfalse

\ifpnas
\documentclass[9pt,twocolumn,twoside,nofootinbib]{pnas-new}
 
    \usepackage{dsfont}

\usepackage{verbatim}
\usepackage{amsmath}
\usepackage{comment}
\usepackage{amsfonts}
\usepackage{amssymb}
\usepackage{numprint}
\usepackage{bm,bbm}
\usepackage{fancyhdr}
\usepackage{empheq}
\usepackage{color} 
\else
  \documentclass[twocolumn,reprint,floatfix,nofootinbib]{revtex4-1}

      \usepackage{dsfont}
    \usepackage{comment}
  \usepackage[utf8]{inputenc}
  \usepackage{newtxtext}
  \RequirePackage[dvipsnames,usenames]{xcolor}
  \usepackage{hyperref}

  \usepackage{color} 
  \usepackage{amsmath} 
  \usepackage{graphicx}

\usepackage{cancel}
\usepackage{url}
\usepackage{xcolor}

\usepackage{bm,bbm}
\usepackage{fancyhdr}
\usepackage{empheq}

\usepackage{graphicx}
\usepackage{amsmath}
\usepackage{amsfonts}
\usepackage{amssymb}
\usepackage{numprint}

\fi

\ifpnas
    \templatetype{pnasresearcharticle} 
 
\else
  \makeatother
 \begin{document}
\fi

\title{Ranking influential nodes in networks from partial information}
\ifpnas

\author[a,b]{Silvia Bartolucci}
\author[a,c]{Fabio Caccioli}
\author[d]{Francesco Caravelli}
\author[e]{Pierpaolo Vivo}

\affil[a]{Dept. of Computer Science, University College London, 66-72 Gower Street WC1E 6EA London (UK)}
\affil[b]{Centre for Financial Technology, Imperial College Business School, South Kensington SW7 2AZ London (UK).} 
\affil[c]{ London Mathematical Laboratory, 8 Margravine Gardens, London WC 8RH (UK).}
\affil[d]{Theoretical Division,
Los Alamos National Laboratory, Los Alamos, New Mexico 87545 (USA)}
\affil[e]{Dept. of Mathematics, King's College London, Strand WC2R 2LS London (UK).}

\else
\author{Silvia Bartolucci$^{1,2}$,  Fabio Caccioli$^{1,4,5}$, Francesco Caravelli$^3$, Pierpaolo Vivo$^{6,\star}$}
\vspace{1cm}
  \affiliation{\\\vspace{0.5cm}$^1$ Dept. of Computer Science, University College London, 66-72 Gower Street WC1E 6EA London (UK).\\ $^2$Centre for Financial Technology, Imperial College Business School, South Kensington SW7 2AZ London (UK). \\ $^3$ Theoretical Division (T4), Condensed Matter \& Complex Systems, Los Alamos National Laboratory, Los Alamos, New Mexico 87545 (USA). \\ $^4$ Systemic Risk Centre, London School of Economics and Political Sciences, WC2A 2AE, London (UK).\\ $^5$ London Mathematical Laboratory, 8 Margravine Gardens, London WC 8RH (UK). \\ $^6$ Dept. of Mathematics, King's College London, Strand WC2R 2LS London (UK).\\
  $^\star$ Corresponding author: {\texttt{pierpaolo.vivo@kcl.ac.uk}}}

\fi

\begin{abstract}

Many complex systems exhibit a natural hierarchy in which elements can be ranked according to a notion of ``influence''. 
While the complete and accurate knowledge of the interactions between constituents is ordinarily required for the computation of nodes' influence, using a low-rank approximation we show that in a variety of contexts local information about the neighborhoods of nodes is enough to reliably estimate how influential they are, without the need to infer or reconstruct the whole map of interactions.
Our framework is successful in approximating with high accuracy different incarnations of influence in systems as diverse as the WWW PageRank, trophic levels of ecosystems, upstreamness of industrial sectors in complex economies, and centrality measures of social networks, as long as the underlying network is not exceedingly sparse. We also discuss the implications of this ``emerging locality'' on the approximate calculation of non-linear network observables.

\end{abstract}

\ifpnas

  \begin{document}

  \maketitle
  \thispagestyle{firststyle}
  \ifthenelse{\boolean{shortarticle}}{\ifthenelse{\boolean{singlecolumn}}{\abscontentformatted}{\abscontent}}{}

\else
  \maketitle

  \section*{Introduction}

\fi

\ifpnas \dropcap{U}nderstanding \else Understanding \fi how complex systems are internally organized and function collectively is a central question in many areas of science, with applications as diverse as hubs detection in epidemiology \cite{epidemiology, bbv-book} and the understanding of  protein-protein interaction networks in biological applications \cite{protein}, to the study of systemic risk in financial networks \cite{bardoscia2021physics,battiston2012debtrank}.  One of the most relevant issues is how constituents -- for example, nodes in a complex networks -- can be efficiently \emph{ranked} according to some meaningful \emph{local} or \emph{global} metric \cite{intro1}. The simplest local observable is the degree $k_i$ of node $i$, which is a useful -- albeit simplistic -- way to assess how ``important'' a node is based only on the number and strength of its immediate contacts. Global measures, such as different types of \emph{centrality}, take a more holistic view of a node in the context of how fast information can travel through it from and towards the outer regions of the network. The downside is that they are usually more difficult to compute accurately, and are very sensitive to minute details of how the network is organized \cite{barabasirank}. Moreover, there is now quite some evidence that many centrality measures are typically correlated with the local degree \cite{corr1,corr2,corr3,corr4,corr5} despite their premise of global character. 

In the context of quintessentially global measures, a prominent way to classify and rank nodes is based on an umbrella notion that we will refer to as  \emph{influence}, which can be defined self-consistently across the whole network. The idea is that a node $i$'s influence $\mathcal{I}_i$ should be slightly larger -- on average -- than that of the nodes ``influenced'' by $i$, i.e. the nodes $j$ in its neighborhood $\partial i$. In formulae,
\begin{equation}
    \mathcal{I}_i = 1 +  \langle \mathcal{I}_{j\in \partial i} \rangle \ .\label{dominance}
\end{equation}
In theoretical ecology, for instance, species forming a complex ecosystem can be assigned a natural ``influence'' (\emph{trophic}) level in the food-chain hierarchy, precisely computed using \eqref{dominance}: apex predators are high-up in the food chain because they either eat top-level predators themselves, or because they consume a large number of medium- or low-level species. The notion of influence as defined in \eqref{dominance} appears however in a much broader range of incarnations: the PageRank algorithm used by Google \cite{Prank} classifies web pages as very relevant if they are linked to by other very relevant addresses, or by very many other pages.  In economics, industrial sectors of different countries can be ranked based on their position and importance in the production chain using so-called \emph{upstreamness} metrics, which measure the distance of a product from final consumption \cite{antras2012econometrica, AntrasFally2012, Fally2012,McNerney2018}. 
 The Katz centrality $\mathcal{K}_i$ of node $i$  -- which is a weighted sum of paths traveling through $i$ \cite{Katz} -- can be itself cast in the form of  \eqref{dominance}, as well as the Mean First Passage Time of a walker from a source to a target node \cite{mfpt0,mfpt,Nicosia}. These examples support the notion that the ``influence'' framework is in principle a rather general and intuitive way to address the ranking problem on networks from a global perspective. 
 
Influence is indeed an intrinsically non-local quantity, which depends on the full set of interactions in the network. Rearranging slightly \eqref{dominance}, we may rewrite the influence vector $\bm{\mathcal{I}}$ of the $N$ nodes as
\begin{equation}
    \bm{\mathcal{I}}= (\mathds{1} - A)^{-1}\bm 1\ ,  
    \label{eq:dominanceA}
\end{equation}
where $\mathds{1}$ is the identity matrix, $\bm 1$ is a vector of all ones, and $A$ is the \emph{interaction matrix}: depending on the circumstances, it may simply be the weighted adjacency matrix of the network (denoted by $\tilde A$ in the following), or a simple function of it that encodes pairwise interactions between constituents. The matrix $G(A)=(\mathds{1} - A)^{-1}$ appearing in \eqref{eq:dominanceA} is referred to as the \emph{resolvent} of $A$. In many interesting examples -- where some form of dissipation or external contribution is involved -- the matrix $A$ is \emph{non-negative} and \emph{sub-stochastic} ($0\leq \sum_j A_{ij}\leq 1$): each entry may represent the fraction of calories a species contributes to the diet of another in an ecosystem, or the fraction of money an industrial sector injects into the economic cycle. 

The seemingly harmless \eqref{eq:dominanceA} has however three important drawbacks: (i) it requires the inversion of a possibly large and ill-conditioned matrix, which makes a numerical
approach computationally intensive and prone to inaccuracies \cite{matrixinversion}, (ii) the nonlinear relation between the influence and the interaction matrix $A$ makes it difficult to infer the functional dependence of the former on network parameters (e.g. the mean degree) from the knowledge of the latter – unless the matrix possess symmetries or a specific structure, and (iii) it requires the complete and accurate knowledge of \emph{all} the pairwise interactions strengths in the network (encoded in $A$), which may be not realistically achieved in many real-life settings.

For instance, reconstructing the food web interaction tables is a notoriously difficult task \cite{dunne}, which can be accomplished with some acceptable accuracy only for very small systems, comprising not more than a handful of species \cite{ecogeneral3}. In economic studies, the compilation of Input/Output tables of developed countries -- unraveling the complex web of interconnections between industrial and financial sectors -- is routinely plagued with all sorts of sampling and surveying errors \cite{KopJansen1994,KopJansen1990}. Calculating the ranking of constituents (e.g., webpages) in large systems is computationally expensive, and the classical PageRank algorithm is difficult to directly implement and optimize in distributed computation frameworks such as Spark \cite{spark}, therefore fostering the quest for faster approximate routes \cite{spark1}. Having a tool to efficiently and accurately rank constituents of a system \emph{we know very little about} would therefore be very useful and of broad applicability.

In this paper, we indeed show that -- in a variety of important contexts -- the influence attributes of a node $i$ in a complex network can be determined with high accuracy using only \emph{local} information. In particular, the detailed knowledge of the complete state of the network is often irrelevant. This \emph{emerging locality} of network influence stems from the observation that a non-negative and sub-stochastic interaction matrix $A$ often displays a ``large'' Perron-Frobenius eigenvalue, which is well-separated from the bulk of all the others. When this happens, $A$ can be faithfully approximated by a rank-$1$ matrix $\hat A$, which retains only some local information stored in $A$.

As long as the network of interactions is ``not too sparse'', this approximation dramatically reduces the amount of detailed information needed to rank constituents as fast as possible, without compromising accuracy. Moreover, it makes it possible to accurately approximate all sorts of non-linear functions of the interaction matrix, as we demonstrate below in the case of exponential and power-law centrality measures.  

\begin{figure}
    \centering
    \includegraphics[width=0.45\textwidth]{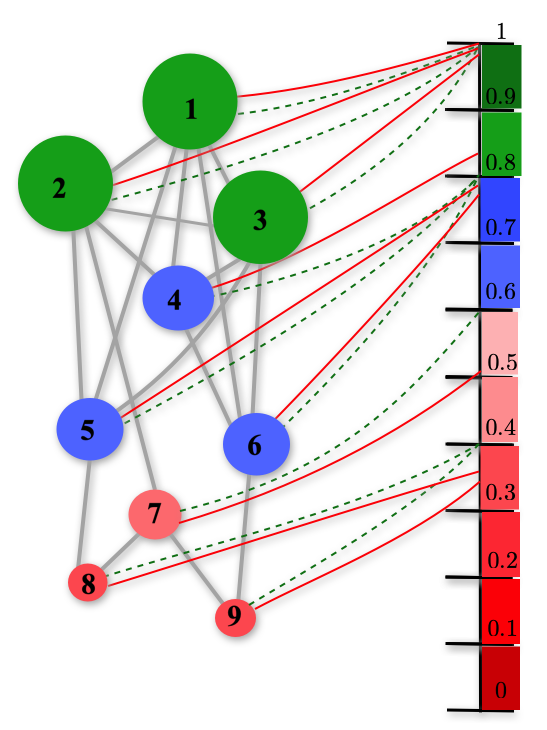}
    \caption{{\bf Illustrative example of the relationship between approximate and exact Katz centrality of nodes.} Comparison between the ranking of influential nodes based on Katz centrality (Eq. \eqref{eq:katzinv} - red solid lines) and our approximate formula (Eq. \eqref{eq:dominancesingle} - green dashed lines) for a small, undirected and dense network of $N=9$ nodes (edges are gray solid lines). The color of each node reflects the Katz centrality ranking (normalized to one), while the node size is proportional to its degree. The matrix $A=\alpha \tilde A=(\eta/k_{\mathrm{max}}) \tilde A$ ($\eta=0.46$) is proportional to the adjacency matrix $\tilde{A}$ of the network, normalized by its largest degree $k_{\mathrm{max}}$. Both the full and approximate centrality measures are normalized to a maximum value of $1$, and compared on the same scale (on the right).  As we can see, the rank-1 formula provides a very good approximation to the exact value of the Katz centrality of each node, and correctly reflects the ``degree centrality'' of each node (represented in the sketch by the node size).  }
    \label{fig:example1}
\end{figure}

\section*{Results}
\subsection*{Rank-1 approximation for influence}

Consider the resolvent $G(A)=(\mathds{1} - A)^{-1}$ of the interaction matrix $A$ of a system of interest, and imagine that a detailed knowledge of all the entries of $A$ is not available. We are only able to accurately estimate $2N$ constants $\bm r=(r_1,\ldots,r_N)$ and $\bm c=(c_1,\ldots,c_N)$, namely the \emph{sums} of the $N$ rows and columns of $A$. This scenario is extremely common: in economics, often only aggregate information is available about sectorial outputs, or interbank exposure \cite{anand}, and the same happens in the study of ecosystems where often only the overall ``energy balance'' of each species can be accurately estimated \cite{berlow}. 

In this scenario, we may construct a simple rank-$1$ approximation $\hat A$ for the matrix $A$ as follows
\begin{equation}
    \hat A=\frac{1}{N}\bm g\bm q^T=
    \begin{pmatrix}
    \frac{g_1 q_1}{N} & \cdots & \frac{g_1 q_N}{N}\\
    \vdots & \ddots & \vdots\\
    \frac{g_Nq_1}{N} & \cdots & \frac{g_Nq_N}{N}
    \end{pmatrix}\ ,\label{avA}
\end{equation}
where the entries of the column vectors $\bm g = (g_1,\ldots,g_N)$ and $\bm q=(q_1,\ldots,q_N)$ can be determined by solving a set of equations imposing the constraint that $A$ and $\hat A$ share the same row and column sums
\begin{align}
r_i= &\sum_j A_{ij}\equiv \frac{\sum_{k} q_k}{N} g_i=\bar q\ g_i\ , \label{eq:fp1}\\
c_j= &\sum_i A_{ij}\equiv \frac{\sum_{k} g_k}{N} q_i=\bar g\ q_j\ . \label{eq:fp2}
\end{align}
This yields eventually the unique matrix (see Supplementary Information)
\begin{equation}
    \hat A =\frac{1}{mN}\bm r\bm c^T\label{hatA}
\end{equation}
with $m=\frac{1}{N} \sum_{ij} A_{ij}=\frac{1}{N}\sum_j c_j=\frac{1}{N}\sum_i r_i$. The rank-$1$ matrix $\hat A$ in \eqref{hatA} is nothing but the Maximum Entropy reconstructed matrix (see e.g. \cite{newcorr2,maxent2}) subject to the row and column constraints in \eqref{eq:fp1} and \eqref{eq:fp2} - see Supplementary Information. In the case when $A$ is the unweighted adjacency matrix of a network, replacing $A$ with its rank-$1$ counterpart $\hat A$ corresponds to the \emph{annealed network} approximation \cite{bianconiannealed,statmech3}. Also, the same result could be interpreted as a first-order expansion of the standard fitness model \cite{statmech1,statmech2}. {{Note that $\hat A$ as constructed here would not necessarily be the optimal (e.g. in the sense of Frobenius norm) rank-$1$ Eckart-Young-Mirsky approximant $\bar A$  \cite{rank1approx1,rank1approx2} of the original matrix $A$. However, the construction of $\bar A$ requires the full and complete knowledge of $A$, which of course would make the whole enterprise pointless.}} 

If the only information we have is about row sums, then the corresponding rank-$1$ matrix is
\begin{equation}
   \hat{A} = \begin{pmatrix}
    \frac{r_1}{N} & \cdots & \frac{r_1}{N}\\
    \vdots & \ddots & \vdots\\
    \frac{r_N}{N} & \cdots & \frac{r_N}{N}
    \end{pmatrix} \label{eq:single}\ .
\end{equation}

Approximating the resolvent - or Green's function - with a low-rank matrix to speed up computations is a customary tool in hard-core condensed matter \cite{lowrankGreen, lowrank2} and fluid dynamics \cite{lowrankfluid}, while being far less common in the context of complexity studies (see however \cite{Lowrank} for a thorough study of the ``low-rank'' hypothesis in complex systems). It is indeed well known \cite{Teukolsky} that matrix inversion scales as $N^3$ in the worst case scenario of a structure-less dense matrix. This implies that for large matrices, the inversion in \eqref{eq:dominanceA} is typically hard.

Clearly, $\hat A$ has a single non-zero, real and positive eigenvalue $\lambda_1=\frac{1}{mN}\sum_j r_j c_j$ due to the Perron-Frobenius theorem, and $N-1$ zero eigenvalues, therefore we may expect that this approximation will work better the larger the ``spectral gap''of the original matrix $A$ is. The spectral gap is defined as $\lambda_1-\max\{|\lambda_2|,\ldots,|\lambda_{N-1}|\}$, with $\lambda_1$ real and $<1$ is the Perron-Frobenius eigenvalue. . Since the spectral gap is usually larger the denser a graph is \cite{gapsparse1,gapsparse2,gapsparse3}, we expect that the approximate formula may not be suitable for ``too sparse'' graphs.   This expectation is in agreement with earlier findings in the context of Mean First Passage Time of random walks on networks \cite{mfpt} and will be extensively tested below. 

In physical terms, the matrix $\hat A$ is a ``perfectly balanced'' version of $A$, where the interactions are spread out as evenly as possible among the constituents: under $\hat A$, every species would consume the same amount of all the others, and every industrial sector would rely on the same fraction of goods produced by all the others to function. 

Armed with this rank-$1$ approximation, we may now proceed to evaluate the approximate resolvent
\begin{equation}
    G(\hat A)=(\mathds{1} - \hat A)^{-1}=
    \mathds{1}+\frac{\hat A}{1-\frac{1}{m N}\sum_j r_j c_j}\ ,\label{approxResolvent}
\end{equation}
using the Sherman-Morrison formula \cite{sherman1950} for the inverse of a rank-$1$ matrix, from which it follows that the influence of the $i$-th node is approximated by 
\begin{equation}
    \mathcal{I}_i \approx 1+\frac{r_i}{1-\frac{1}{m N}\sum_j r_j c_j}\ \label{eq:dominance}
\end{equation}
(or
\begin{equation}
    \mathcal{I}_i \approx 1+\frac{r_i}{1-\frac{1}{ N}\sum_j r_j }\ \label{eq:dominancesingle}
\end{equation}
in the case of row-constraints only, a formula that we used in \cite{BCCV} to study production networks, using techniques from disordered systems).
Within our rank-$1$ approximation, the influence of node $i$ is fully determined by the interplay of (i) \emph{local} information, namely the sum of incoming weights into node $i$, and (ii) a suitable average of incoming weights across \emph{all} nodes in the network. In cases where the interaction matrix $A$ is proportional to the adjacency matrix -- e.g. in the case of Katz centrality -- the $\{r_i\}$ and $\{c_i\}$ are obviously proportional to the out-strength and in-strength of node $i$, respectively. {The out-strength $k_i^{(out)}$ of node $i$ is equal to $\sum_j \tilde A_{ij}$, whereas the in-strength $k_i^{(in)}$ is equal to $\sum_j \tilde A_{ji}$. If the network is unweighted, the strengths reduce to the out-degree and in-degree, respectively.} 
Our formula \eqref{eq:dominance} therefore explicitly shows {how \emph{global} quantities (in this case the Katz centrality of node $i$) and \emph{local} ones (in this case its weighted degree) can be strongly correlated (being in fact proportional to each other)}, a fact that has been previously noted in several network instances \cite{corr3,evans,newcorr,newcorr2} {(see Fig. \ref{fig:example1} for a sketch of the mutual relation between Katz centrality, our approximate formula, and degree centrality)}. 

We further corroborate this observation by studying a class of network transformations (\emph{Maslov-Sneppen} \cite{maslovsneppen1,maslovsneppen2}) that preserve the degree sequence while drastically altering the topology, and we show that indeed the Katz centrality of nodes is preserved in the not-too-sparse regime.

It is interesting to compare the rank-$1$ approximation $G(\hat A)$ for the resolvent $G(A)$ in \eqref{approxResolvent} with an alternative approach, which is possibly the simplest approximation/reconstruction scheme one could adopt. The resolvent $G(A)$ can be formally expanded in a power series as
\begin{equation}
    G(A) = \mathds{1}+A+A^2+\ldots\ ,
\end{equation}
which admits an interesting interpretation e.g., in the economical context: the first contribution accounts for the direct increase in output of all industrial sectors that is necessary to
meet an increase in final demand. The second contribution accounts for the increase in output
that is needed to meet the increment in input required by all sectors to meet the increase in
final demand. This chain of $k$-th order effects is encoded in the $k$-th term of the expansion, and unravels the technological interdependence of the productive system within an economy. The crudest approximation to the resolvent consists in truncating the above expansion at a given finite order. As we show in the Supplementary Information, however, these approximations do not reproduce the influence rankings as accurately as our approximate formula Eq. \eqref{eq:dominance}.

In the next section, we spell out in more detail a few commonly used incarnations of the influence relation \eqref{eq:dominanceA}, before putting the approximate formula \eqref{eq:dominance} to the test on synthetic and empirical networks.

\subsection*{Notable examples of influence}
In social networks, the influence of nodes can be assessed via the \emph{Katz centrality} \cite{Katz}, which takes into account the total number of walks between a pair of agents. 
The Katz centrality computes the relative influence of a node within a network by measuring the number of its immediate neighbors (first degree nodes) and also all other nodes in the network that connect to the node under consideration through these immediate neighbors. Connections made with distant neighbors are, however, penalized by an attenuation factor $\alpha<1$.  In formulae, the Katz centrality of the $i$-th node of a network of size $N$ described by the $(0,1)$-adjacency matrix $\tilde A$ is given by
\begin{equation}
   \mathcal{K}_i=\sum_{\ell=1}^\infty\sum_{j=1}^N\alpha^\ell (\tilde A^\ell)_{ji} \ .\label{Katz1}
\end{equation}
It measures the total number of paths of given length reaching $i$ from any other node, where paths of length $\ell$ are given a discounted weight by a factor $\alpha^\ell$. Eq. \eqref{Katz1} can be cast in the ``influence'' form \eqref{eq:dominanceA} as
\begin{equation}
\bm{\mathcal{K}}(A) =(\mathds{1} - A)^{-1}\bm 1 - \bm 1\ ,
    \label{eq:katzinv}
\end{equation}
with $A=\alpha\tilde A$.\\

 Similarly, the PageRank algorithm used by Google \cite{Prank} returns rankings $\bm{\mathcal{R}}$ of $N$ websites based on the assumption that more ``important'' websites are likely to receive more links from other websites. Given a web environment composed of $N$ sites, one forms the re-scaled adjacency matrix $\tilde A$ is such a way that $\tilde A_{ij}$ is the ratio between number of links pointing from page $j$ to page $i$ to the total number of outbound links of page $j$. Then one defines the \emph{damping factor} $d$ as the probability that an imaginary surfer who is randomly clicking on links will keep doing so at any given step (typically, $d\simeq 0.85$). Given these ingredients, the ranking of pages follows the formula 
\begin{equation}
   \bm{\mathcal{R}}(A)=\frac{1-d}{N}(\mathds{1}-A)^{-1}\bm 1\ ,
   \label{eq:prank}
\end{equation}
where $A=d\tilde A$.

Influence measures can also be introduced in the context of hierarchies in complex ecosystems. In an ecosystem composed of $N$ species, one introduces the interaction matrix $S_{ij}$ that represents the fraction of biomass transferred from the species $j$ to the species $i$, often described as the fraction of $j$ in the ``diet'' of $i$. The \emph{trophic levels} of species are the relative positions that they occupy in the ecosystem \cite{hannon,trophic1,trophic2,dunne,allesina1,allesina2}, i.e., apex predators have a larger trophic level than phytoplankton. 
The trophic level of species $i$ is defined as
\begin{equation}
    \mathcal{T}_i=1+\sum_{j} S_{ij} \mathcal{T}_j\ ,
\end{equation}
or if expressed in the typical ``influence'' form
\begin{equation}
    \bm{\mathcal{T}}(A) =(\mathds{1}-A)^{-1}\bm{1}\ ,
    \label{eq:trophic}
\end{equation}
with $A\equiv S$. This definition indeed implies that $\mathcal{T}_i=1$ -- the lowest possible value -- is reserved to species $i$ such that $A_{ij}=0$ for all $j$, i.e. those who occupy the lowest level of the food chain. Trophic levels have been studied in ecology to analyze structural properties of ecosystems \cite{trophic1,trophic2, trophic3, price,allesina1,allesina2,john,Trophicbio,Trophicnetworkbio2}, more recently the same concept has been applied to investigate and classify financial strategies \cite{trophicfinance}, to quantify amplification of shocks in financial exposure networks \cite{mackaysystemic}, as well as functional and dynamical properties of nodes in networks \cite{TrophicNetwork1,mackay}, including the effect of global directionality in directed networks \cite{GlobalDirectionality}.

Finally, in the context of complex economies and so-called input-output models \cite{Leontiefbook}, measures of influence are used to determine the role and importance of industrial sectors within an economy \cite{BCCV}. We can define the so-called \emph{upstreamness} \cite{antras2012econometrica, AntrasFally2012, Fally2012} as follows:

\begin{equation}
    \bm{\mathcal{L}}(A)=(\mathds{1}-A)^{-1}\bm 1\ ,\label{upstream}
\end{equation}
where the matrix $A_{ij}$ is the dollar amount of sector $i$'s output that is needed to produce one dollar of $j$'s output. The upstreamness of an industry, or financial sector, is used to assess  their relative position in the value production chain \cite{Fally2012}, and their relation to economic growth \cite{McNerney2018}.

We are now ready to put the approximate, \emph{local} formulae \eqref{eq:dominance} and \eqref{eq:dominancesingle} for each node's influence to the test: in the next section, we assess how well the formulae perform on network synthetic data (in the context of Katz centrality), as well as on different incarnations of the ``influence'' framework in the context of economics, social networks, relevance of web pages, and ecology -- as described above.

\subsection*{Test of the formula on data}
In this section, we will test how the rank-$1$ formula with single \eqref{eq:dominancesingle} and double constraints \eqref{eq:dominance} performs on synthetic network data and on six sources of empirical data, considering different incarnations of the influence ranking. Tests include computing the Katz centrality on synthetic Erd\H{o}s-R\'enyi (ER) and Scale Free (SF) networks, measuring the so-called \emph{upstreamness}, i.e., inter-sectorial dependencies in complex economies,  \emph{Katz centrality} of users of social networks, relevance of web-pages using \emph{PageRank}, and \emph{trophic levels} of species within an ecosystem.

\begin{figure}
    \centering
    \includegraphics[scale=0.33]{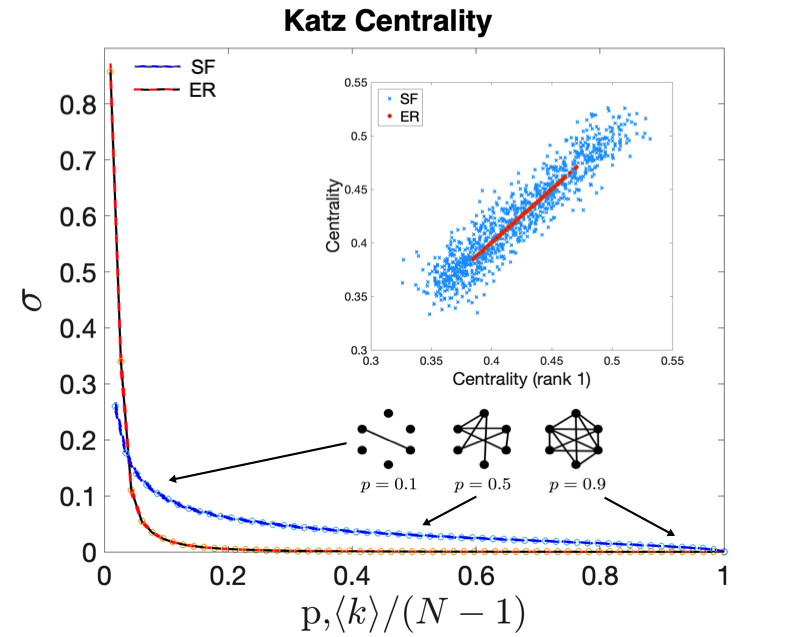}
    \caption{{\bf Katz centrality of Erd\H{o}s-R\'enyi and Scale-Free networks.} We plot the $\sigma$ metric (defined in \eqref{eq:sigmametric}) as a function of the sparsity parameter $p$ for ER, and the equivalent $\langle k\rangle/(N-1)$ for SF networks $(\gamma = 2.2)$ of size $N=1000$, where $\langle k\rangle$ is the mean degree. The thickness of the red and blue curves corresponds to the spread of $\sigma$ across $M=3600$ instances, whereas the black dashed curves correspond to the average $\overline{\sigma}$ over such  instances.  In the inset, we provide instead a scatter plot of numerical vs. approximate centrality measures for $p=0.5$.}
    \label{fig:comparison}
\end{figure}

\textit{Katz centrality of synthetic networks.} We first test  the rank-$1$ formula on synthetic data. We consider two types of networks in Fig. 
\ref{fig:comparison}: (i) directed Erd\H{o}s-R\'enyi (ER) random graphs, with $p$ being the probability of having an edge between any two nodes of the graph and (ii) scale-free (SF) graphs with a power-law degree distribution. 

We use $\alpha=0.3/\bar d$, where $\bar d$ is the maximum eigenvalue. For Erd\H{o}s-R\'enyi graphs, $p=\langle k\rangle/(N-1)$, with $\langle k \rangle$ the mean degree, which can be also used as a measure of sparsity for Scale Free graphs. A scale free graph is defined by the degree distribution $P(k)\propto k^{-\gamma}$ (here we use  $\gamma=2.2$), and can be constructed using the Bianconi-Barab\'{a}si algorithm \cite{barabasi1999emergence,barabasibianconi}. We define the following metric

\begin{equation}
    \sigma = \Big\langle\bigg|\frac{\mathcal{K}_i^{(\mathrm{num})}}{\mathcal{K}_i^{\mathrm{(approx)}}}-1\bigg|\Big\rangle\ ,
    \label{eq:sigmametric}
\end{equation}
where $\mathcal{K}_i^{(\mathrm{num})}$ is the Katz centrality of the $i$-th node of the $m$-th instance $(m=1,\ldots,M)$, evaluated by numerical inversion \eqref{eq:katzinv}, $\mathcal{K}_i^{(\mathrm{approx})}$ is the approximate Katz centrality of the same node specializing the rank-$1$ formula \eqref{eq:dominance}, and the average  $\langle \cdot \rangle$ is performed over the $M$ instances and the $N$ nodes. In Fig. \ref{fig:comparison}, we plot $\sigma$ as a function of $p$ for ER and SF graphs averaged over $M=3600$ instances, where a lower $\sigma$ clearly indicates a better agreement between the exact and approximate formulae.

We see that for ER graphs, the agreement between the rank-$1$ approximation of the Katz ranking and the actual value evaluated via matrix inversion is almost perfect, unless the graph is very sparse. For scale-free graphs the agreement is still good, but the noise is again larger for sparser networks. This is in agreement with a similar observation in the context of Mean First Passage Time for random walks on networks \cite{mfpt} that the accuracy of the approximate formula \eqref{eq:dominance} depends on the spectral gap of the interaction matrix, which makes it not suitable for ``too sparse'' networks.

\begin{figure}
    \centering
    \includegraphics[scale=0.3]{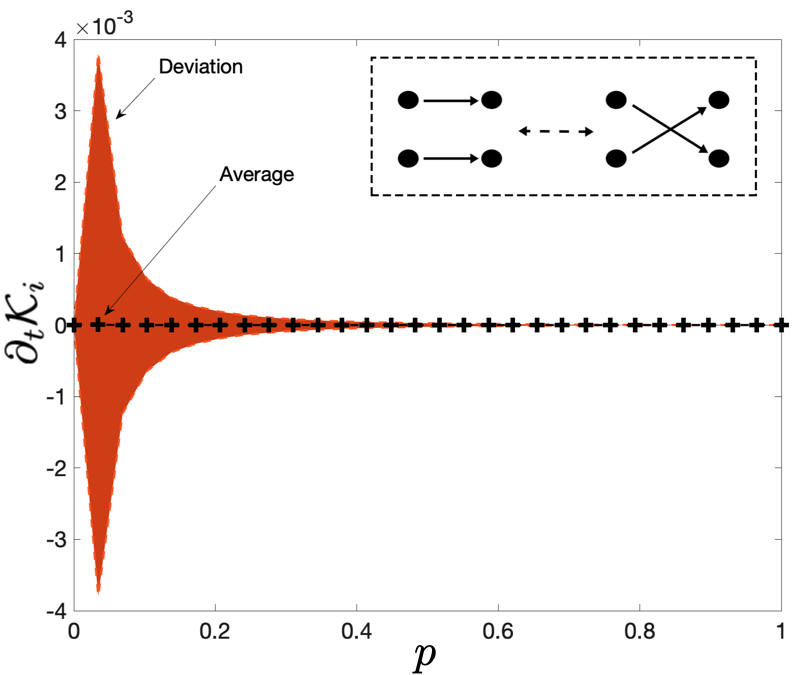}
    \caption{{\bf Maslov-Sneppen dynamics.} Black crosses: Average value of the derivative of the Katz centrality ranking ($\alpha=0.3$). The average is over $100$ initially drawn ER networks with $N=100$, and the total number of Maslov-Sneppen steps ($T=100$), and it is plotted  as a function of the edge probability $p$. The red area is the standard deviation of $\partial_t \mathcal{K}_i^{(\ell)}$ across time/instances. }
    \label{fig:maslovsneppen}
\end{figure}

We now test the fact that the Katz centrality of a node in not-too-sparse networks should be largely determined by the node's strength (or degree) by considering a class of simple degree-preserving transformations that dramatically alter the topology of a network. The \emph{Maslov-Sneppen} rule \cite{maslovsneppen1,maslovsneppen2}, shown in Fig. \ref{fig:maslovsneppen}, works as follows: given two pairs of nodes $i$ and $j$ and $i^\prime$ and $j^\prime$ of a directed graph,  such that the links $i\rightarrow j$ and $i^\prime\rightarrow j^\prime$ do exist, one performs the rewiring $i\rightarrow j^\prime$ and $i^\prime \rightarrow j$. Clearly, both the out-strengths of $i$ and $i^\prime$ are unchanged, and so are the in-strength of $j$ and $j^\prime$.  We generate $M=100$ Erd\H{o}s-R\'{e}nyi graphs of size $N=100$ with edge probability $p$, and perform 100 dynamical (time) steps $t$. At each time step, we select two pairs of edges at random, and we rewire them, as depicted in the inset of Fig. \ref{fig:maslovsneppen}. For each network $\ell$ in the sample, we calculate the Katz centrality $\mathcal{K}_i^{(\ell)}(t)$ ($\alpha=0.3$) of each node $i$. Then, as  a function of $p$ and for every sample $\ell$ and every node, we calculate the discrete time derivative $\partial_t \mathcal{K}_i^{(\ell)}(t)=\mathcal{K}_i^{(\ell)}(t)-\mathcal{K}_i^{(\ell)}(t-1)$. Then, we calculate $\langle \partial_t \mathcal{K}_i^{(\ell)}(t)\rangle=\frac{1}{NM(T-1)} \sum_{t=1}^{T-1} \sum_{i=1}^N \sum_{\ell=1}^M  \partial_t \mathcal{K}_i^{(\ell)}(t)$, which is the average change in the centrality value across nodes, time, and samples. This average clearly depends only on the edge probability $p$. In Fig. \ref{fig:maslovsneppen}, we see that
the average derivative is close to zero across the full range of $p$, while the fluctuations -- assessed by the standard deviation across instances and time -- also become negligible after $p\approx 0.3$. This confirms that the Katz centrality is strongly correlated to the degree sequence of the network, and even more so the denser the network is.

\begin{figure*}[htb!]
\centering
\includegraphics[width=.95\textwidth]{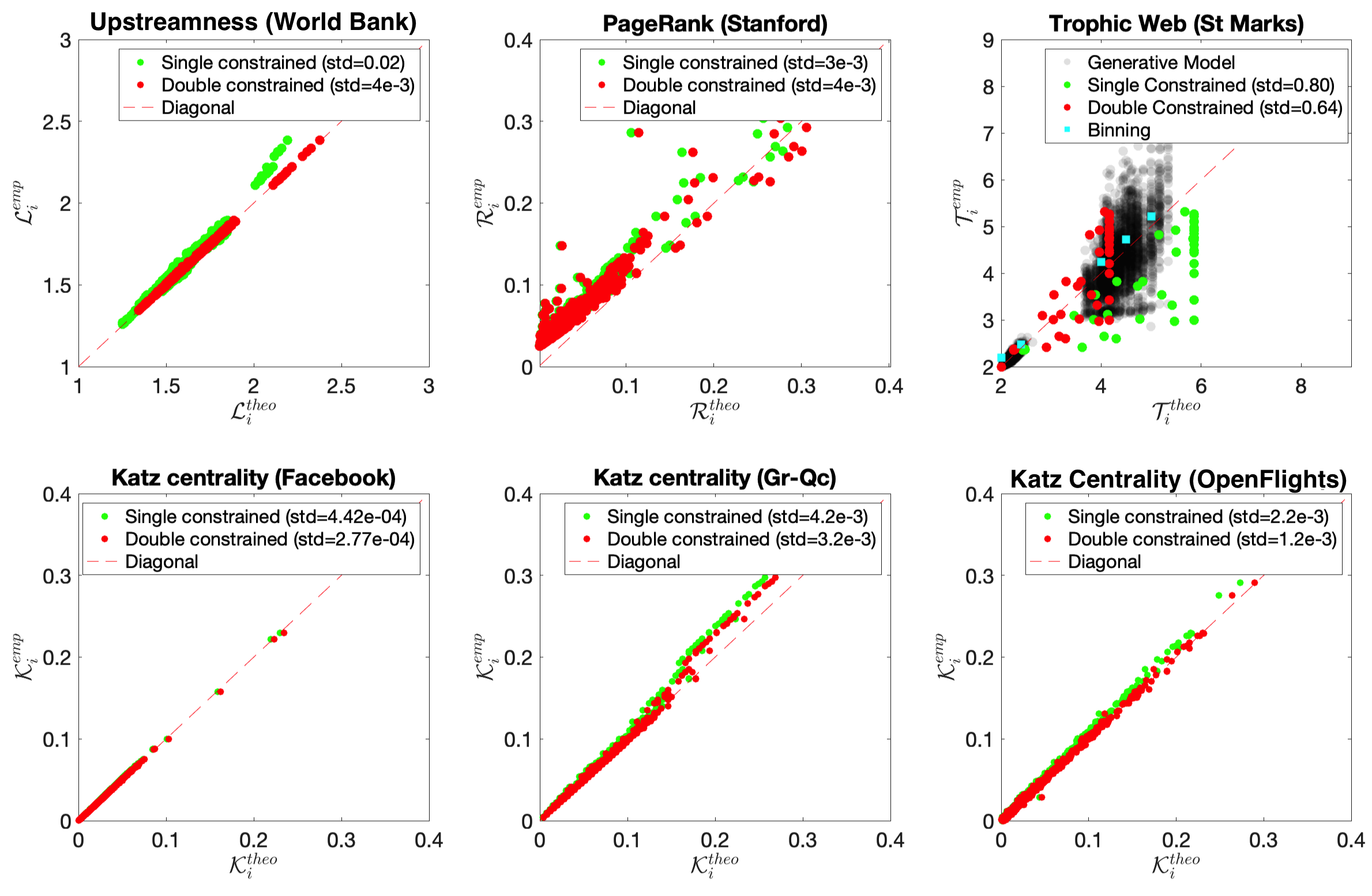}\\
\caption{{\bf Test of the approximate formula for influence on real data. }Scatter plots of {different incarnations of} the \emph{empirical} influence {(computed from the data interaction matrices using Eq. \eqref{eq:dominanceA}}) vs. {our approximate formulae, for respectively} the single-{[green dots]} (Eq. \eqref{eq:dominancesingle}) and doubly-constrained {[red dots]} (Eq. \eqref{eq:dominance}) models. In the top left figure we compute the upstreamness of industrial sectors in complex economies for the World Bank dataset; in the top center figure, the PageRank of webpages for the 4000 nodes Stanford dataset;  in the top right figure we compute the trophic levels of species in the St. Marks ecosystem. In the bottom row, from left to right we compute in order the Katz centrality of the Facebook users dataset, the arXiv Gr-Qc collaboration network and the OpenFlights dataset (details on the datasets are provided in the main text).}
\label{figure}
 \end{figure*}

\textit{Katz centrality of social network data.} The Katz centrality \eqref{Katz1} is also calculated on the widely used Ego-Facebook dataset \cite{McAuley}, comprising the matrix of interactions of $4039$ users (nodes) and 176468 friendships (edges). We chose the value of the damping factor $\alpha=2.87\times 10^{-4}$, which is the largest admissible value for which (i) the series in \eqref{Katz1} converges, and (ii) the matrix $\alpha\tilde A$ is sub-stochastic. {We also evaluated the Katz centrality on the arXiv GR-QC collaboration network, composed of 5242 nodes and 14490 edges, representing co-authorship on General Relativity and Quantum Cosmology papers \cite{networkdata1}; we also consider the OpenFlights network dataset \cite{networkdata1,networkdata2}, a directed graph composed of 2939 nodes (representing airports) and 30501 edges (weighted by the number of air routes connecting them).}
Each point in the lower panels of Fig. \ref{figure} represents the centrality of one of the nodes. We compare the empirical Katz centrality against the theoretical rank-1 approximation with single (green dots, see \eqref{eq:dominancesingle}) and double (red dots, see \eqref{eq:dominance}) constraints.

\textit{Economic dataset.} The upstreamness \eqref{upstream} is computed using the data from the National Input-Ouptut Tables (NIOT) of the World Input Output Database (WIOD - 2013 release) \cite{Timmer}. The NIOT dataset represents the flow of money between $35$ industrial sectors for $39$ world economies, and it includes the years $1995-2011$. The full list of countries and economic sectors as well as a scheme of the structure of the input-output tables can be found in \cite{Timmer,BCCV}.

The matrix $A$ from which we calculate the upstreamness is computed from the full input-output table of each country normalising the entries of the I-O table by the total output (the row sum of the matrix plus the demand of external entities) \cite{BCCV}. Each point in the upper left panel of Fig. \ref{figure} represents the average upstreamness (over all sectors) of a given country in a given year (the plot thus includes $39 {\rm (countries)}\times 17{\rm (years)}=663$ data points).

\textit{PageRank.}
The PageRank \eqref{eq:prank} is evaluated on the \textit{wb-cs.Stanford} dataset, a collection of $9914$ indexed web-pages of the World Wide Web \cite{stanford,Hirai} dating from 2001, for a value of the damping parameter equal to $d=0.3$ (upper middle panel, Fig. \ref{figure}).

\textit{Trophic Levels.} The trophic levels (defined in \eqref{eq:trophic}) are computed for the $51$ species that are part of the St. Marks food web \cite{Baird} (upper right panel, Fig. \ref{figure}). Since on average the size of ecosystems does not normally exceed $\sim 200$ species, to match the empirical data with the rank-1 approximation we have also constructed the corresponding stochastic niche model in order to generate and reproduce the typical trophic diameter of the food web, as it is customary in this type of analysis \cite{ecogeneral1,ecogeneral2,ecogeneral3,ecogeneral4}.
We used a ``cascade'', or niche model, extracted from the empirical food web $A$, not only reproducing the mean and the variance of the empirical data, but rather estimating the full distribution \cite{ecosystembook}. Each black dot in the upper right panel of Fig. \ref{figure} corresponds to one of the $T_k(A_s)$ values $(k=1,\ldots,51$ and $s=1,\ldots,100)$ generated by $100$ instances of the random model with $N=51$ generative species each -- plotted against the corresponding theoretical value estimated using \eqref{eq:dominance} and \eqref{eq:dominancesingle}. While there are sample to sample fluctuations, binning (purple dots in Fig. \ref{figure} - upper right panel) reveals that the average trophic levels nicely follow the theoretical approximations. More generally, we observe an excellent agreement between our approximate formula -- which does not require any matrix inversion, nor the full details of the pairwise interaction matrix -- and the empirical data in all cases, especially using the doubly-constrained version of the influence, Eq. \eqref{eq:dominance}.
{For all datasets, we measured the Pearson's and Spearman's correlation coefficients between the empirical values and those estimated using the approximate formula. We observe correlation coefficients well above 95\% for all cases--in fact above 98\% in most instances--with the exception of the St. Marks food web, for which we observe correlations around 0.6 (Spearman's coefficient) and 0.8 (Pearson's coefficient). This further supports the claim that the approximate formula provides quite good estimates of nodes influence in a variety of contexts.}
\\

\subsection*{Effect of sparsity on the accuracy of the local approximation}

{In this section, we perform numerical experiments on synthetic data to assess the performance of the local approximation for the influence (Eq. \eqref{eq:dominancesingle}).}

We start from a random $N\times N$ dense matrix $A_N(t=0)$ (typically, $N=1000$) with exponentially distributed entries (representing a \emph{fully connected} and weighted initial network) and collect the row sums $\{r_1,\ldots,r_N\}$. Then, at each discrete time $t=1,2,\ldots$ we pick one nonzero entry ($(A_N)_{k\ell}$) at random and set it to zero (unless it was the last nonzero entry of its row). Then, we redistribute the nullified $(A_N)_{k\ell}$ entry evenly to all other nonzero entries of its row $k$ -- to preserve each $r_k$ along the process.

This way, we progressively ``sparsify'' -- in a controlled way -- the matrix $A_N(t=0)$ while retaining its initial row sums unchanged. The process ends when only one nonzero entry per row (clearly equal to $r_k$ in row $k$) survives.

For each matrix $A_N(t)$ produced with this procedure, we compute the ``real'' influence values $\bm{\mathcal{I}}(t)=(\mathds{1}-A_N(t))^{-1}\bm 1$, {which we compare with the approximate values $\hat{\bm{\mathcal{I}}}$ from Eq. \eqref{eq:dominancesingle}. We repeat the above steps over several realizations of the matrix $A_N(t=0)$ and the sparsification trajectory, and we compute over time as a measure of the estimation error the quantity 
\begin{equation}
    \sigma (t) = \left\langle\left|\frac{\mathcal{I}_i(t)}{\hat{\mathcal{I}}_i}-1\right|\right\rangle,
\end{equation}
where the average $\langle \cdot \rangle$ is taken over nodes and realizations.}

{In the numerical experiment we consider in this section, one instance of $A_N(t=0)$ is generated as follows: We first create a matrix  $W$ of size $N\times(N+1)$, where all entries are drawn from an exponential distribution of average $1$. We then set $A_N(t=0)_{ij}=W_{ij}/\sum_{k=1}^{N+1} W_{ik}$, with $i$ and $j$ both ranging between $1$ and $N$. In this way, we obtain an $N\times N$ full sub-stochastic matrix (we did set diagonal entries to zero as we do not consider the case of a node in the network pointing to itself).}
 \begin{figure}
     \centering
     \includegraphics[width=0.45\textwidth]{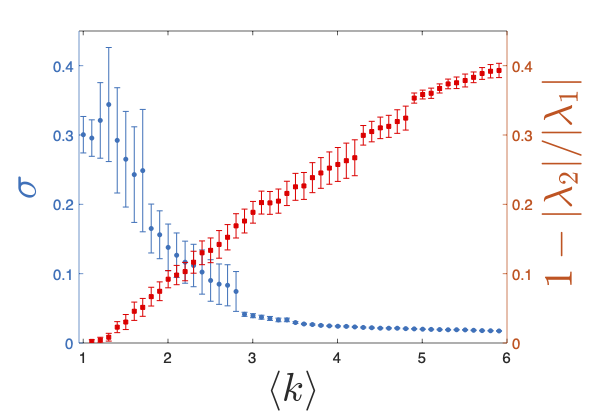}
     \caption{{\bf Sparsification procedure.} {Left axis: Estimation error associated with the sparsification procedure as a function of the average degree of the network obtained through the sparsification procedure. Averages are computing over $1000$ realizations for networks of size $N=1000$. Right axis: spectral gap as a function of the average degree computed over $30$ realizations of networks of size $N=1000$. Bars refer to standard errors. The larger the average degree, the smaller the estimation error and the larger the spectral gap.}}
     \label{fig:sparsification}
 \end{figure}
 
 We report in figure \ref{fig:sparsification} results obtained averaging over $1000$ instances of networks of size $N=1000$.
{In the $x$-axis of the figure, we report the average degree of the network during the different steps of the sparsification procedure, so that time increases moving from the right to the left along the $x$-axis.
We observe that the estimation error increases over time as the network becomes sparser, but that it remains fairly low throughout the sparsification procedure -- the error remains roughly below $5\%$ up until the initially fully connected network has been reduced to a network of average degree $\langle k\rangle\approx 3$.}

{In figure \ref{fig:sparsification}, we also report the spectral gap associated with the networks obtained during the sparsification procedure. As a measure of spectral gap, we consider the quantity $g=1-|\lambda_2|/|\lambda_1|$,
where $\lambda_1$ and $\lambda_2$ are the eigenvalues with largest and second largest modulus respectively.  
}
{Interestingly, and as expected, we find that the behavior of $\sigma$ is in direct correspondence with the decrease over time in the spectral gap that follows from the graph becoming progressively sparser, compatibly with the findings in \cite{mfpt}.}

These findings confirm that our approximate formula \eqref{eq:dominancesingle} provides a very good approximation to the true influence values up to an already sparse regime characterized by a large percentage of links removed.

\subsection*{Beyond influence}
The rank-$1$ approximation for the resolvent allows us to get analytical access to a number of other network observables that can be written as functions $F(A)$ of the interaction matrix $A$.
The key identity is the Dunford-Taylor formula \cite{matrixf}, with which we can write a matrix function as the following integral 
\begin{equation}
F(A)=\frac{1}{2\pi \mathrm{i}} \oint_\Gamma dz (z\mathds{1}-A)^{-1} F(z)\ ,\label{eq:dunford}
\end{equation}
assuming that $F$ is analytic, and  the domain  $\Gamma$ encircles the spectrum of $A$.

Approximating $A$ with $\hat A$ as in \eqref{avA} and replacing the resolvent $G(A)$ with $G(\hat A)$ as defined in \eqref{approxResolvent}, we obtain the following approximation (see Supplementary Information for the derivation)
\begin{equation}
    F(A)\approx F(0) \mathds{1}+\frac{F(\xi)-F(0)}{mN\xi}\bm r\bm c^T\ ,\label{DunfTaylorFA}
\end{equation}
where $ \xi=\sum_i r_ic_i/(mN)$.

For instance, many graph invariants, such as topological indices, can be written in the form of a matrix exponential. A non-local quantity of interest is the so-called \emph{graph communicability} \cite{estradabook} given by
\begin{equation}
c_{ij}=({\rm e}^{\beta \tilde A})_{ij}
\label{commexp}
\end{equation}
where ${\rm e}^{\beta \tilde A}=\sum_{k=0}^\infty \beta^k \tilde A^k/k!$ is the  exponential of the adjacency matrix. This measure accounts for all possible channels of communication between node $i$ and $j$, giving more weight to  the shorter paths connecting them on the network represented by $\tilde A$. 
The communicability of a node $i$ is given  by $c_i =\sum_{j} c_{ij}$, and the total communicability of the graph is defined as the {\em Estrada index} $EE(\tilde A)=\sum_i c_i$.
Using \eqref{DunfTaylorFA} we can also provide a natural explanation for the observed correlation between the strengths of nodes (or simply their degrees for unweighted graphs) and the communicability of the network, observed in \cite{comm4}. In particular, we can approximate the communicability matrix as follows
\begin{equation}
   {\rm e}^{\beta \tilde A}\approx \mathds{1}+\frac{\exp\left(\beta\frac{\bm k^{(out)}\cdot \bm k^{(in)}}{\sum_j k_j^{(out)}}\right)-1}{\bm k^{(out)}\cdot \bm k^{(in)}}\bm k^{(out)}\bm k^{(in)T}\ ,
    \label{eq:explain}
\end{equation}
where $\bm k^{(out)}$ and $\bm k^{(in)}$ are the column vectors of out- and in-strengths of the network, and $\cdot$ denotes the dot product. For uncorrelated networks, the average value of the communicability over an ensemble of networks can therefore be computed -- at least in principle -- using
\begin{equation}
    \langle f(\bm k^{(in)},\bm k^{(out)})\rangle=\sum_{\bm m,\bm n}f(\bm m,\bm n)\prod_{j=1}^N p(m_j,n_j)\ ,
\end{equation}
where $p(k^{(in)},k^{(out)})$ is the joint distribution of in- and out-strengths of the nodes.

\begin{figure}
    \centering
    \includegraphics[width=0.47\textwidth]{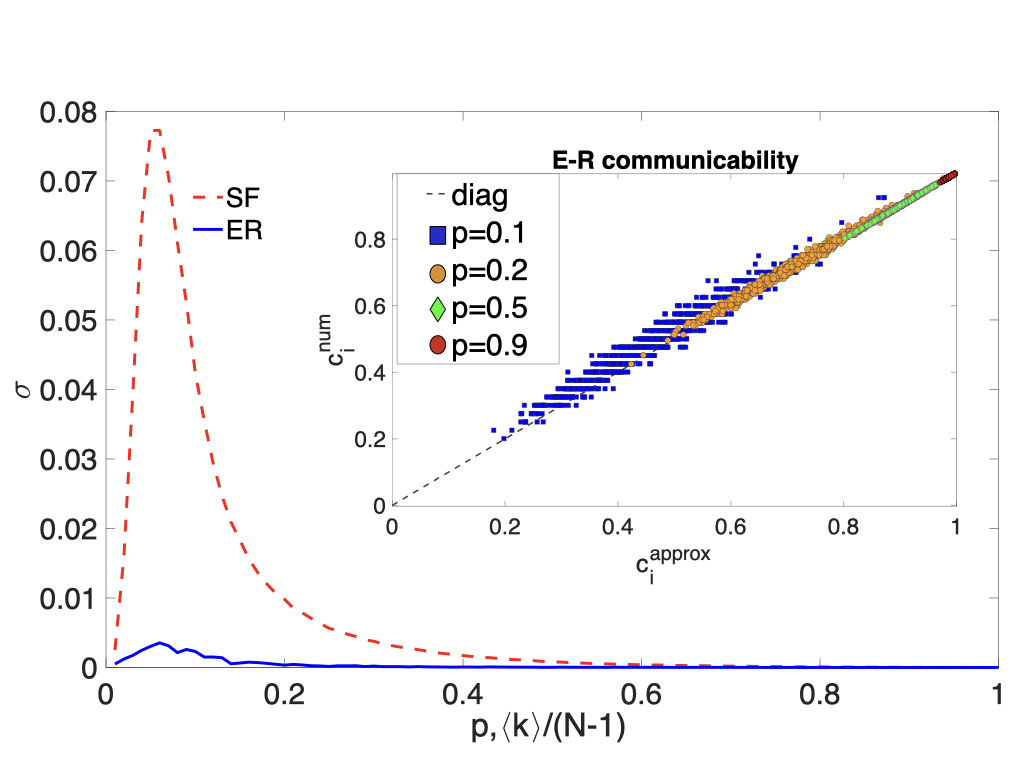}
    \caption{{\bf Communicability of Erd\H{o}s-R\'enyi and Scale-Free networks. }Plot of the $\sigma$-metric for the communicability of Erd\H{o}s-R\'enyi \eqref{eq:sigmametric2} and Scale Free graphs as a function of the edge probability $p$ and $\langle k\rangle/(N-1)$ for $\gamma=2.2$ and $\beta=0.3$. The thickness of the red curve corresponds to the spread of $\sigma$ across $M=100$ instances, whereas the black dashed curve corresponds to the average $\overline{\sigma}$ over such instances. \textit{Inset:} Communicability $c_i$ of the $i$-th node of Erd\H{o}s-R\'enyi network computed using our approximate formula \eqref{eq:explain} against $c_i^{(num)}$ obtained from Eq. \eqref{commexp} for different edge probabilities $p$ ($N=1000, \beta=0.3$).}
    \label{fig:communicability}
\end{figure}

 In order to see how well our approximate formula performs as a function of the connectivity of the graph, we define the metric (analog of \eqref{eq:sigmametric})
\begin{equation}
    \sigma = \Big\langle \bigg |\frac{c_i^{(\mathrm{num})}}{c_i^{\mathrm{(approx)}}}-1\bigg|\Big\rangle\ ,
    \label{eq:sigmametric2}
\end{equation}
where $c_i^{(\mathrm{num})}$ is the communicability of the $i$-th node of the $m$-th instance $(m=1,\ldots,M)$ of an Erd\H{o}s-R\'enyi graph of size $N=100$, evaluated using \eqref{commexp}, while  $c_i^{(\mathrm{approx})}$ is the approximate communicability of the same node obtained from \eqref{eq:explain}. The average  $\langle \cdot \rangle$ is performed over the nodes of an instance. In Fig. \ref{fig:communicability}, we plot $\sigma$ as a function of the edge probability $p$, where a lower $\sigma$ clearly indicates a better agreement between the exact and approximate formulae. The thickness of the red curve corresponds to the spread of $\sigma$ across $M=100$ instances, whereas the black dashed curve corresponds to the average $\overline{\sigma}$ over such instances. As expected, the denser the graph the better the agreement with our rank-1 approximation. In the inset of Fig. \ref{fig:communicability} we show a scatter plot of the communicability of the $i$-th node computed using our approximate formula \eqref{eq:explain} against empirical data for different edge probabilities $p$.

These results can be easily generalized to further incarnations of exponential measures. These includes for example the generalized Estrada index defined as
$GE(\tilde A)=\text{Tr}({\rm e}^{\beta \tilde A})=\sum_i \lambda_i({\rm e}^{\beta \tilde A}) $, where $\text{Tr}$ is the matrix trace, originally introduced to measure the degree of folding of a protein and later extended to characterize topological properties of graphs \cite{estrada,estradabook,comm1,comm2,comm3,matching}.

Evaluating these types of observable with high accuracy requires non-trivial computational resources for large graphs, as they normally scales as $N^3$ with the number of nodes \cite{comm2,comm3}.  In addition, there are not many exact or even approximate formulae in the case of dense graphs, exception made for very specific topologies for which the graph eigenvalues are exactly known. Results for dense \textit{weighted} graphs are even scarcer. In this context, accurate formulae derived from a simple rank-$1$ approximation may be potentially quite useful.

Our rank-1 approximation using the Dunford-Taylor identity \eqref{DunfTaylorFA} can also be applied to further nonlinear observables in the context of centrality measures.
A straightforward extension is for example the application to centrality measures of the form $F(A)=A^k$. Using \eqref{DunfTaylorFA} we obtain  $\langle A^k\rangle_{ij}\approx\frac{\xi^{k-1}}{mN}(\bm r\bm c^T)_{ij}$. A notable example for this class of centrality measure is the so-called diffusion centrality \cite{diffcentr,generalcentrality}, which is defined in terms of a weighted sum of matrix powers and used in simple models of information diffusion to measure the expected number of times a node receives a given piece of information.

\section*{Discussion}\label{sec:discussion}

Constituents of an interacting system organized in a network structure may contribute very differently to the functional tasks the organization is set to achieve as a whole. It is therefore very important to devise meaningful ways to rank them according to useful metrics. Inherently \emph{local} rankings -- based e.g. on the degree of each node -- are easy and quick to compile, but only provide limited information. \emph{Global} rankings -- e.g. centrality measures -- take instead into account the full state of the network and provide in principle a more complete picture, but are often cumbersome to compute, and highly sensitive to the accuracy with which the full set of pairwise interactions are known. 

We showed that a large class of nominally \emph{global} rankings given by the ``influence'' levels $\mathcal{I}_i$ (see \eqref{eq:dominanceA}) may be accurately estimated using only \emph{local} information about incoming weights, and may not require the full knowledge -- or even an accurate reconstruction -- of the whole set of pairwise interactions. The accuracy of the approximation depends on the existence of a ``large'' spectral gap in the interaction matrix: we found that this condition is not too hard to materialize in empirical data sets -- as we show in  Fig. \ref{figure} -- at least away from the ``high sparsity'' regime.

Our findings follow from the observation that the matrix $\hat A$ that best approximates a given interaction matrix $A$ (in the Maximum Entropy sense), and shares with $A$ the row/column sums, has rank-$1$. This provides an immediate route -- via the Sherman-Morrison formula \cite{sherman1950} -- to approximate the resolvent matrix $G(A)=(\mathds{1}-A)^{-1}$, which appears in the general definition of ``influence'' (see \eqref{eq:dominanceA}) and thus in all its particular incarnations described above. The \emph{emerging locality} that stems from this approximation -- and its {empirical} validity across fields and disparate applications -- is particularly appealing for various reasons:
\begin{itemize}
    \item In ecology/economy -- among many other fields -- a complete and accurate control over the interactions between species/productive sectors is plagued with a wealth of theoretical and practical problems. Our approach offers an avenue to forego the full knowledge of the interaction matrix and determine the set of influence levels with high accuracy using only limited and local information about the activity of each player.
    \item Centrality measures are known to correlate with the degree of the corresponding nodes. At least in the context of ``spectrally gapped'' networks, our approach makes this connection clear and transparent, providing an exact expression and clear interpretation for the proportionality constant, and allowing for more straightforward analytical treatments.
    
\item Being the resolvent $G(A)$ the building block of any analytic matrix function via the Dunford-Taylor formula \eqref{eq:dunford}, the rank-$1$ approximation $G(\hat A)$ for the resolvent actually carries over to a number of non-linear network observables beyond the influence. We have presented examples for exponential and power-law centrality measures, and provided evidence that our rank-$1$ approximation can be very accurate, and more analytically tractable than its original version, at least in the not-too-sparse regime.
\end{itemize}

A few comments on our framework are in order:
\begin{enumerate}
    \item The theoretical results and empirical validation presented here are limited to quantities (such as the Katz centrality) that can be cast in the ``influence'' form of Eq. \eqref{eq:dominanceA}, and do not necessarily extend beyond this class.
    \item Our approximate formulae based on a rank-$1$ approximation are less and less accurate the sparser the network is. This, indeed, may appear to be a severe limitation in practice, as many empirical networks are known or believed to be very sparse. As a matter of fact, there is a growing interest in the study of dense networks (both empirical and synthetic), with many interesting applications  \cite{Bianconidense,PREdense} -- for which our framework is perfectly applicable. Moreover, our experiments with sparser networks demonstrate that our approximate formula remains valid for ``moderately'' sparse graphs, and -- even when it does not nail each of the $N$ values of the influence very accurately -- it still provides a meaningful relative ranking of constituents in a fast, efficient, and easily interpretable way.  
    \item If we only have at our disposal the sums of rows and/or columns of the interaction matrix (as we assumed throughout this paper), it is of course impossible to know \emph{a priori} what its spectral gap (or any other meaningful measure of sparsity) is -- therefore the expected accuracy of our formula cannot be easily predicted. It is therefore fair to say that the optimal setting in which to employ our formula is when there is some information about the sparsity level of the network under consideration, in addition to the row/column sums information. 
\end{enumerate}
In future studies, it will be important to (i) characterize in a more quantitative way the accuracy of the approximation as a function of the spectral gap, (ii) study more thoroughly the effect of graph heterogeneity on the quality of the approximation, and (iii) explore rank-$k$ (with $k>1$) systematic approximations for the resolvent.

\newcommand\ackcontent{The work of Francesco Caravelli was carried out under the auspices of the NNSA of the U.S. DoE at LANL under Contract No. DE-AC52-06NA25396. Francesco Caravelli was also financed via DOE LDRD grant PRD20190195.}

\ifpnas
\showmatmethods{} 

\acknow{\ackcontent}
\showacknow{} 

\else 
\section*{Acknowledgments}
\ackcontent

\fi

\section*{Author Contributions}
All authors designed the study, derived the analytical results, run the numerical experiments, wrote and reviewed the manuscript.
\section*{Competing Interests Statement}
The authors declare no competing interests.
\section*{Data availability section}
All raw data analyzed are available via the public repositories cited in the bibliography. The code used to produce the figures and the derived data is available upon request from the corresponding author.

\section*{Supplementary Information}

\subsection*{Uniqueness of rank-$1$ approximation}
We show here that the rank-$1$ matrix $\hat A$ in \eqref{avA} with prescribed row and column sums ($\bm r$ and $\bm c$) is unique. Since $\hat A$ is rank-$1$, we must have $\bm g=\lambda \bm r$ and $\bm q=\mu \bm c$. Imposing the conditions on the row and column sums (see e.g. \eqref{eq:fp1}), we are unable to fix the constants $\lambda$ and $\mu$ separately, but their product $\lambda\mu$ must be equal to $1/m$. Computing $\hat A = (1/N)\bm g\bm q^T=\lambda\mu (1/N)\bm r\bm c^T=(1/mN)\bm r\bm c^T$ as in \eqref{hatA}.

\subsection*{Truncation}

\begin{figure}
    \centering
    \includegraphics[scale=0.15]{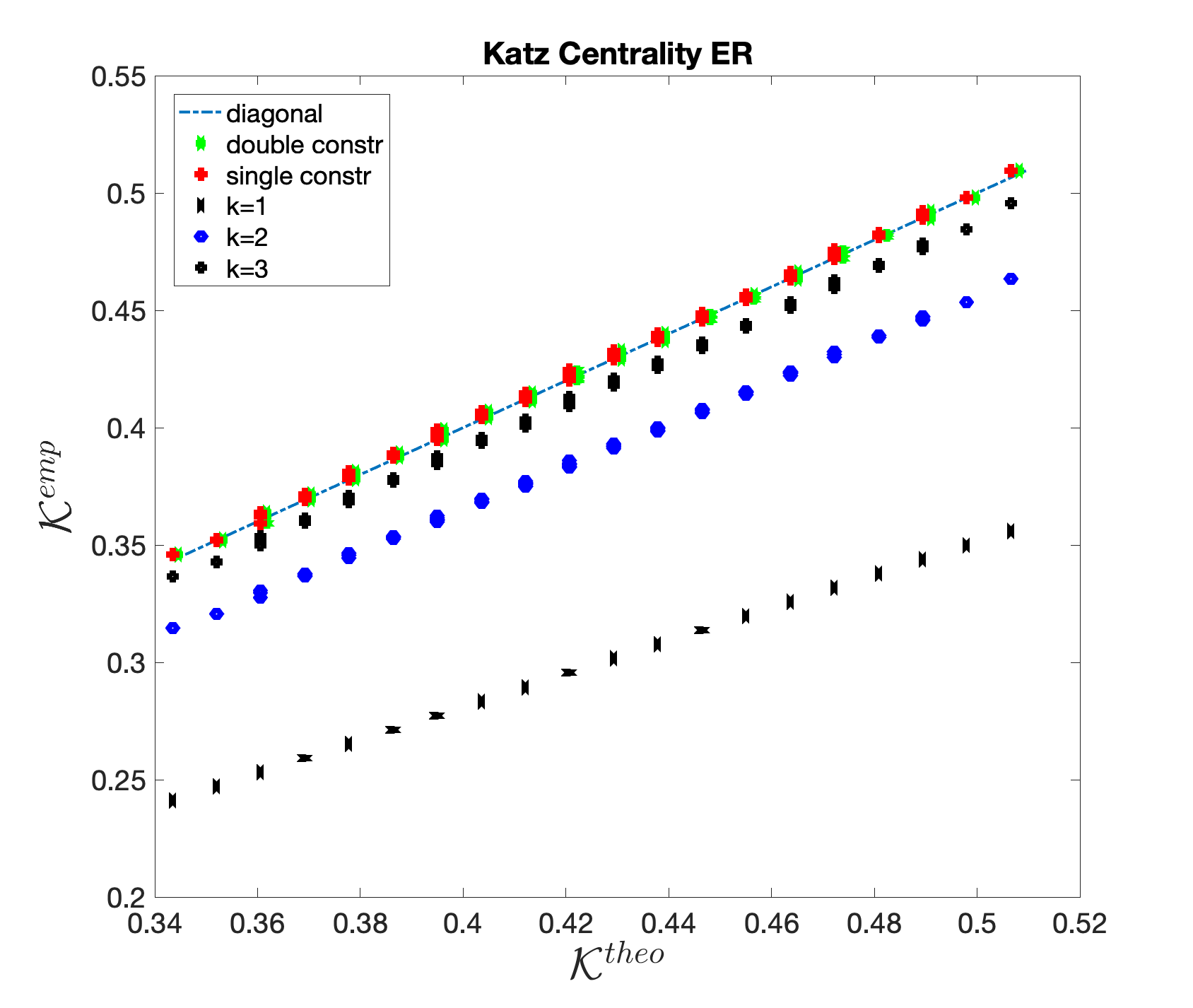}
    \caption{{\bf Different approximations of Katz centrality. }For a single instance of an unweighted and undirected ER graph with edge probability $p=0.5$ and $N=100$ nodes, we provide the scatter plot of the Katz centrality $\mathcal{K}_i$ (see \eqref{Katz1} for $\alpha=0.3$) of a selection of nodes, obtained from a $k$-truncation -- i.e. replacing $G(A)$ with $G_k(A)$ from \eqref{Gk} in \eqref{eq:katzinv} -- together with our single- and doubly-constrained approximation (\eqref{eq:dominance} and \eqref{eq:dominancesingle}).}
    \label{fig:comparisonKtruncation}
\end{figure}

{In Fig. \ref{fig:comparisonKtruncation}, we show that the approximation for the resolvent $G(A)$ consisting in truncating its series expansion at the $k$-th term is usually inferior to the rank-$1$ formula \eqref{approxResolvent} in ``spectrally-gapped'' cases \cite{mfpt}, unless a large number of terms is retained. Moreover, a direct construction of the ``best'' $k$-th order approximant:
\begin{equation}
    G_k(A)=\mathds{1}+\sum_{\ell=1}^k A^\ell\label{Gk}
\end{equation}
starting from the knowledge of row and column sums alone is not an obvious task. As we see, only the infinite $k$ limit reproduces our formula accurately.}

\subsection*{$\hat A$ can be written as the Maximum Entropy matrix with prescribed row and column sums} 
Giving the entries $0\leq \hat A_{ij}\leq 1$ a probabilistic interpretation, we ask what the assignment of $\hat A_{ij}$ should be that maximizes the entropy
\begin{equation}
    S = -\sum_{i,j}\hat A_{ij}\ln \hat A_{ij}
\end{equation}
subject to the constraints $r_i=\sum_j \hat A_{ij}$ and $c_j =\sum_i \hat A_{ij}$. Constructing the Lagrangian
\begin{align}
 \nonumber   \mathcal{L} &=\sum_{i,j}\hat A_{ij}\ln \hat A_{ij}-\lambda_0\left(\sum_{i,j}\hat A_{ij}-Nm\right)\\
 &-\sum_i\lambda_i\left(\sum_j \hat A_{ij}-r_i\right)-\sum_j\mu_j\left(\sum_i \hat A_{ij}-c_j\right)
\end{align}
and computing its stationary points, we obtain the equations

\begin{equation}
   \ln\hat A_{ij}+1-\lambda_0-\lambda_i-\mu_j=0\Rightarrow \hat A_{ij}=\frac{1}{Z}\phi_i\psi_j\label{maxenthatA}
\end{equation}   
where we set $1/Z=\exp(\lambda_0-1)$, $\exp(\lambda_i)=\phi_i$ and $\exp(\mu_j)=\psi_j$. Therefore, we see that the MaxEnt matrix $\hat A$ with prescribed row and column sums is rank-$1$. Imposing the row and column sums constraints (or simply appealing to uniqueness), we finally obtain again $\hat A = (mN)^{-1}\bm r\bm c^T$.

\vspace{10pt}
\subsection*{Generalized  nonlinear formula}
First, we use the following representation of the matrix $F(A)$
\begin{align}
    F(A)&=\frac{1}{2\pi \mathrm{i}}\oint_\Lambda dz F(z) (z\mathds{1}-A)^{-1} \nonumber \\
    &=\frac{1}{2\pi \mathrm{i}}\oint_\Lambda dz \frac{F(z)}{z} \left(\mathds{1}-\frac{1}{z}A\right)^{-1}\ ,
\end{align}
where the integration is over a curve $\Lambda$ in the complex plane, which contains all eigenvalues of $A$.

We now apply the approximate expression for the resolvent \eqref{approxResolvent},
\begin{equation}
    F(A) \approx\frac{1}{2\pi \mathrm{i}}\oint_\Lambda dz \frac{F(z)}{z} \left(\mathds{1}+\frac{1}{mNz-\sum_{i=1}^N r_ic_i}\bm r\bm c^T \right)\ ,
\end{equation}
where we recall the definitions $r_i=\sum_{j} A_{ij}$, $c_j=\sum_{i} A_{ij}$, and $m=\sum_i r_i=\sum_j c_j$. It is easy to show that $ \xi=\sum_i r_ic_i/(mN)$ is included within the curve $\Lambda$. Therefore, using

\begin{equation}
\frac{1}{z(z-\xi)}=\frac{1}{ \xi}\left(\frac{1}{z- \xi}-\frac{1}{z} \right)\ ,
\end{equation}
and applying Cauchy's residue theorem to the singular points $z=\xi$ and $z=0$, we obtain the identities 
\begin{align}
\frac{1}{2\pi \mathrm{i}}\oint_\Lambda dz \frac{F(z)}{z}&=F(0)\ ,\\ 
\frac{1}{2\pi \mathrm{i}}\oint_\Lambda dz \frac{F(z)}{z(z-\xi)}
&=\frac{F(\xi)}{\xi}-\frac{F(0)}{\xi}\ .
\end{align}

We thus obtain the formula \eqref{DunfTaylorFA} of the main text

\begin{equation}
    F(A)\approx F(0) \mathds{1}+\frac{F(\xi)-F(0)}{mN\xi}\bm r\bm c^T\ .
\end{equation}

\end{document}